\documentclass{article}
\usepackage{graphicx}
\usepackage{here}
\begin{document}
\def\beq{\begin{equation}}
\def\eeq{\end{equation}}
\def\bea{\begin{eqnarray}}
\def\eea{\end{eqnarray}}
\def\bq{\begin{quote}}
\def\eq{\end{quote}}
\def\ve{\vert}
\def\nnb{\nonumber}
\def\ga{\left(}
\def\dr{\right)}
\def\aga{\left\{}
\def\adr{\right\}}
\def\lb{\lbrack}
\def\rb{\rbrack}
\def\rar{\rightarrow}
\def\lrar{\Longrightarrow}
\def\llar{\Longleftarrow}
\def\nnb{\nonumber}
\def\la{\langle}
\def\ra{\rangle}
\def\nin{\noindent}
\def\ba{\begin{array}}
\def\ea{\end{array}}
\def\bm{\overline{m}}
\def\ind{\indexentry}
\def\c{\clubsuit}
\def\s{\spadesuit}
\def\b{\bullet}
\def\als{\alpha_s}
\def\as{\ga\frac{\bar{\alpha_s}}{\pi}\dr}
\def\asr{\ga\frac{{\alpha_s}}{\pi}\dr}
\def\gg{ \la\alpha_s G^2 \ra}
\def\ggg{g^3f_{abc}\la G^aG^bG^c \ra}
\def\gggg{\la\als^2G^4\ra}
\def\lnu{\log{-\frac{q^2}{\nu^2}}}
\begin{flushright}
PM 01/55
\end{flushright}
\vspace*{1cm}
\begin{center}
\section*{  
 Status of the {\boldmath$B^0_{(s)}$-$\bar B^0_{(s)}$} mixing\\ 
from QCD spectral sum Rules \footnote{Talk presented at the 1st High-Energy
Physics Madagascar International Conference Series (HEP-MAD'01), 27th sept-5th Oct. 2001,
Antananarivo.}}

\vspace*{1.5cm}
{\bf Stephan Narison} \\
\vspace{0.3cm}
Laboratoire de Physique Math\'ematique\\
Universit\'e de Montpellier II\\
Place Eug\`ene Bataillon\\
34095 - Montpellier Cedex 05, France\\
Email: qcd@lpm.univ-montp2.fr\\
\vspace*{1.5cm}
{\bf Abstract} \\ \end{center}
\vspace*{2mm}
\noindent
{
In this talk, I present new results  \cite{NOMURA} obtained from QCD spectral
sum rules (QSSR), on the bag constant parameters entering in the
analysis of the $B^0_{(s)}$-$\bar B^0_{(s)}$ mass-differences.  
Taking the average of the results from the Laplace and moment sum rules, one obtains to order $\alpha_s$:
$f_B\sqrt{\hat B_B}\simeq (228 \pm 61)~{\rm MeV}~,
{f_{B_s}\sqrt{\hat B_{B_s}}}/{f_{B}\sqrt{\hat B_{B}}}\simeq
1.18\pm 0.03$, in units where $f_\pi=130.7$ MeV.
 Combined with the experimental data
on the mass-differences
$\Delta M_{d,s}$, one obtains the constraint on the CKM weak mixing angle:
$|V_{ts}/V_{td}|^2\geq  20.2(1.3)$. Alternatively, using the weak mixing angle from the
analysis of the unitarity triangle and the data on $\Delta M_d$, one predicts
$\Delta M_s=18.3(2.1)~ps^{-1}$ in agreement with the present experimental lower bound and within
the reach of Tevatron 2. }

\section{Introduction}
\nin
 $ B^0_{(s)}$ and $\bar{B}^0_{(s)}$ are not eigenstates of the weak hamiltonian, such that their
oscillation frequency is gouverned by their mass-difference $\Delta M_q$. The measurement 
by the UA1 collaboration \cite{UA1} of a large value of $\Delta M_d$ was the {\it first}
indication of a heavy top-quark mass.
In the SM, the mass-difference
is approximately given by \cite{BURAS}:
\bea\label{deltam}
\Delta M_q&\simeq& \frac{G^2_F}{4\pi^2}M^2_W|V_{tq}V^*_{tb}|^2
S_0\ga\frac{m^2_t}{M^2_W}\dr\eta_BC_B(\nu)\frac{1}{2M_{B_q}}
\la \bar{B}^0_q|{\cal O}_q(\nu)|B^0_q\ra~
\eea
where the $\Delta B=2$ local operator ${\cal O}_q$ is defined as:
\beq\label{eq:operator}
{\cal O}_q(x)\equiv (\bar b\gamma_\mu L q)(\bar b\gamma_\mu L q)~,
\eeq
with: $L\equiv (1-\gamma_5)/2$ and $~q\equiv
d,s,$;
$S_0,~\eta_B$ and $C_B(\nu)$ are short distance quantities and Wilson coefficients which are
calculable perturbatively, while the matrix element $\la \bar{B}^0_q|{\cal O}_q|B^0_q\ra$ requires
non-perturbative QCD calculations, and is usually parametrized for $SU(N)_c$ colours as:
\beq
\la \bar{B}^0_q|{\cal O}_q|B^0_q\ra=N_c\ga 1+\frac{1}{N_c}\dr
f^2_{B_q}M^2_{B_q}B_{B_q}~. 
\eeq
$f_{B_q} $ is the $B_q$ decay constant normalized as $f_\pi=92.4$ MeV, and
$B_{B_q}$ is the so-called bag parameter which is $B_{B_q}\simeq 1$ if one uses a vacuum
saturation of the matrix element.
From Eq. (\ref{deltam}), it is clear that the measurement of $\Delta
M_d$ provides the one of the CKM mixing angle $|V_{td}|$ if one uses $|V_{tb}|\simeq 1$. One can
also extract this quantity from the ratio:
\bea\label{mass}
\frac{\Delta M_s}{\Delta M_d}&=&\Big{\vert}\frac{V_{ts}}{V_{td}}\Big{\vert}^2\frac{M_{B_d}}{M_{B_s}}
{\frac{\la \bar{B}^0_s|{\cal O}_s|B^0_s\ra}{\la \bar{B}^0_d|{\cal
O}_d|B^0_d\ra}}
\equiv \Big{\vert}\frac{V_{ts}}{V_{td}}\Big{\vert}^2\frac{M_{B_d}}{M_{B_s}}\xi^2~,
\eea
since in the SM with three generations and unitarity constraints, $|V_{ts}|\simeq |V_{cb}|$. Here:
\beq\label{bbs}
\xi\equiv
\sqrt{\frac{g_s}{g_d}}\equiv\frac{f_{B_s}\sqrt{B_{B_s}}}{f_{B}\sqrt{B_{B}}}~.
\eeq
The 
great advantage of Eq. (\ref{mass}) compared with the former relation in Eq. (\ref{deltam}) is
that in the ratio, different systematics in the evaluation of the matrix element tends to cancel
out, thus providing a more accurate prediction. However, unlike $\Delta M_d= 0.473(17) 
~ps^{-1}$, which is measured with a good precision \cite{PDG}, the determination of
$\Delta M_s$ is an experimental challenge due to the rapid oscillation of the $B^0_s$-$\bar{B}^0_s$
system. At present, only a lower bound of 13.1 $ps^{-1}$ is
available at the 95\% CL from experiments
\cite{PDG}, but this bound already provides a strong constraint on
$|V_{td}|$.
\section{Two-point function sum rule} 
\nin
Ref.  \cite{PICH} has extended the analysis
 of the
$K^0$-$\bar{K}^0$ systems of \cite{RAFAEL}, using two-point correlator of the four-quark operators into
the analysis of the quantity
$f_{B}\sqrt{B_{B}}$ which gouverns the
$ B^0$-$\bar{B}^0$ mass difference. The two-point correlator defined as:
\beq\label{twopoint}
\psi_H(q^2) \equiv i \int d^4x ~e^{iqx} \
\la 0\vert {\cal T}
{\cal O}_q(x)
\ga {\cal O}_q(0)\dr ^\dagger \vert 0 \ra ~,
\eeq
is built from the $\Delta B=2$ weak operator defined in Eq. (\ref{eq:operator}). The corresponding
Laplace (resp. moment) sum rules are:
\beq
{\cal L}(\tau)=\int_{4M^2_B}^{\infty}dt~{\rm e}^{-t\tau}{\rm Im}\psi_H(t)~,~~~~~~~~~
{\cal M}_n=\int_{4M^2_B}^{\infty}{dt~ t^{n}}~{\rm Im}\psi_H(t)~,~~~~~~~~~
\eeq
 The two-point function approach is very convenient due to
its simple analytic properties which is not the case of approach  based on three-point functions
\footnote{For detailed criticisms, see \cite{SNB}.}. However, it involves non-trivial QCD
calculations which become technically complicated when one includes the contributions of radiative
corrections due to non-factorizable diagrams. These perturbative radiative corrections 
due to {\it factorizable and non-factorizable} diagrams  have been already
computed in \cite{PIVO} (referred as NP), where it has been found that the factorizable corrections
are large while the non-factorizable ones are negligibly small. NP analysis has confirmed the
estimate in \cite{PICH} from lowest order calculations, where under some assumptions on the
contributions of higher mass resonances to the spectral function, the value of the bag parameter
$B_B$ has been found to be:
\beq\label{eq:pivo1}
B_{B_d}(4m_b^2)\simeq (1\pm 0.15)~.
\eeq
This value is comparable with the one $B_{B_d}= 1$ from the
vacuum saturation estimate, which is expected to be a quite good approximation due to the relative
high-scale of the $B$-meson mass. 
Equivalently, the corresponding RGI quantity is:
\beq\label{eq:pivo2}
\hat B_{B_d}\simeq (1.5\pm 0.2)
\eeq
where we have used the relation:
\bea
 B_{B_q}(\nu)={\hat
B_{B_q}}{\alpha_s^{-\frac{\gamma_0}{\beta_1}}}\aga 1-
\ga \frac{5165}{12696}\dr\ga\frac{\alpha_s}{\pi}\dr\adr,
\eea
with $\gamma_0=1$ is the anomalous dimension  of the
operator
${\cal O}_q$ and $\beta_1=-23/6$ for 5 flavours. The NLO corrections have been obtained in the
$\overline{MS}$ scheme \cite{BURAS}. We have also used, to this order, the value \cite{SNMAS,SNB}:
\beq\label{runmass}
\bar{m}_b(m_b)=(4.24\pm 0.06)~{\rm GeV}~,
\eeq
and $\Lambda_5=(250\pm 50)~{\rm MeV}$ \cite{BETHKE}.
In a forthcoming paper \cite{NOMURA}, we study {\it (for the first time)}, from the QCD spectral
sum rules (QSSR) method, the
$SU(3)$ breaking effects on the ratio:
$
\xi$
defined previously in Eq. (\ref{bbs}),
where a similar analysis of the ratios of the decay constants has given the values \cite{SNFBS}:
\beq\label{fbs}
\frac{f_{D_s}}{f_{D}}\simeq 1.15\pm 0.04~,~~~
\frac{f_{B_s}}{f_{B}}\simeq
1.16\pm 0.04~.
\eeq
We also improve the previous result on $B_{B_d}$ by the inclusion of the $B^*-B^*$ resonances into
the spectral function. 
\section{Results and implications on \boldmath $|V_{ts}/V_{td}|^2$ and 
\boldmath$\Delta M_s$}
\nin
We deduce by taking the average from the moments and Laplace sum rules
results \cite{NOMURA}:
\beq\label{xi}
\xi \equiv\frac{f_{B_s}\sqrt{B_{B_s}}}{f_{B}\sqrt{B_{B}}}\simeq 1.18\pm 0.03~,~~~~
f_B\sqrt{\hat B_B}\simeq (228 \pm 61)~{\rm MeV},
\eeq
in units where $f_\pi=130.7$ MeV.
As expected, we have smaller errors for the ratio $\xi$ due to the
cancellation of the systematics, while for
$f_B\sqrt{\hat B_B}$, the error comes mainly and equally from the pole mass $M_b$ and
the truncation of the PT series where we have estimated the strength of the $\alpha_s^2$ contribution
assuming a geometric growth of the PT coefficients.
These results can be compared  with different lattice and phenomenological determinations given in
\cite{LATT,LATT2}. By comparing the ratio with the one of $f_{B_s}/f_{B_d}$ in Eq. (\ref{fbs})
\footnote{One can notice that similar strengths of the $SU(3)$ breakings have been obtained
for the $B\rar K^*\gamma$ and $B\rar K l\nu$ form factors \cite{SNFORM}.}, one can conclude
(to a good approximation) that:
\beq
\hat{B}_{B_s}\approx \hat{B}_{B_d}\simeq (1.41\pm 0.33)\lrar B_{B_{d,s}}(4m^2_b)\simeq (0.94\pm
0.22)~,
\eeq
indicating a negligible $SU(3)$ breaking for the bag parameter. For a consistency, we have used the
estimate to order $\alpha_s$ \cite{SNFB1}:
\beq
f_B\simeq (1.47\pm 0.10)f_\pi~,
\eeq 
and we have assumed that the error from $f_B$ compensates the one in Eq. (\ref{xi}). The result is
in excellent agreement with the previous result of \cite{PIVO} in Eqs (\ref{eq:pivo1}) and
(\ref{eq:pivo2}). 
 Using the experimental values:
\bea
\Delta M_d&=& 0.472(17)~ps^{-1}~,~~~~~~~~~
\Delta M_s\geq 13.1~ps^{-1}~~(95\%~{\rm CL}),
\eea
one can deduce from Eq. (\ref{mass}):
\beq
\rho_{sd}\equiv\Big{\vert}\frac{V_{ts}}{V_{td}}\Big{\vert}^2\geq 20.2(1.3).
\eeq
Alternatively, using: $\rho_{sd}\simeq 28.4(2.9)$ obtained by using the Wolfenstein parameters 
determined in
\cite{LATT}, we deduce:
\beq
\Delta M_s\simeq 18.4(2.1)~ps^{-1}~,
\eeq
in good agreement with the present experimental lower bound and within the reach of Tevatron run 2
experiments.
\section{Conclusions}
\nin
We have applied QCD spectral sum rules for extracting {\it (for the first time)}
the $SU(3)$
breaking parameter in Eq. (\ref{xi}). 
The phenomenological consequences of our results for the $B^0_{d,s}$-$\bar B^0_{d,s}$ mass-differences
and CKM mixing angle have been discussed. An
extension of this work to the study of the
$B^0_{s,d}-\bar B^0_{s,d}$ width difference is in progress \cite{PREP}.
\section*{Acknowledgements} It is a pleasure to thank my collaborators Kaoru Hagiwara and Daisuke Nomura.

\end{document}